\documentclass[12pt]{emulateapj}
\usepackage[]{natbib}
\usepackage{graphicx, amsmath}
\def\fesc{{$f_{\rm esc}^{{\rm Ly}\alpha}$}}

\begin{document}

\title{{\it HST\/} Emission Line Galaxies at $z \sim 2$:  The Ly$\alpha$
Escape Fraction} 
\shorttitle{The $z \sim 2$ Ly$\alpha$ Escape Fraction}
\author{Robin Ciardullo\altaffilmark{1}, 
Gregory R. Zeimann\altaffilmark{1}, 
Caryl Gronwall\altaffilmark{1}, 
Henry Gebhardt\altaffilmark{1}, 
Donald P. Schneider\altaffilmark{1},
Alex Hagen\altaffilmark{1},
A.I. Malz\altaffilmark{1}}
\affil{Department of Astronomy \& Astrophysics, The Pennsylvania
State University, University Park, PA 16802}
\email{rbc@astro.psu.edu}

\author{Guillermo A. Blanc}
\affil{Observatories of the Carnegie Institution for Science,
Pasadena, CA, USA}

\author{Gary J. Hill}
\affil{McDonald Observatory, The University of Texas at Austin,
Austin, TX 78712}

\author{Niv Drory}
\affil{Department of Astronomy, The University of Texas at
Austin, 2515 Speedway, Stop C1400, Austin, TX 78712, USA}

\and

\author{Eric Gawiser}
\affil{Department of Physics and Astronomy, Rutgers, The State
University of New Jersey, Piscataway, NJ 08854, USA}

\altaffiltext{1}{Institute for Gravitation and the Cosmos, The Pennsylvania 
State University, University Park, PA 16802}

\clearpage

\begin{abstract}
We compare the H$\beta$ line strengths of $1.90 < z < 2.35$ star-forming
galaxies observed with the near-IR grism of the {\sl Hubble Space
Telescope\/} with ground-based measurements of Ly$\alpha$ from the
HETDEX Pilot Survey and narrow-band imaging.  By examining the line ratios
of 73 galaxies, we show that most star-forming systems at this epoch have a 
Ly$\alpha$ escape fraction below $\sim 6\%$.  We confirm this result by 
using stellar reddening to estimate the effective logarithmic extinction of
the H$\beta$ emission line ($c_{{\rm H}\beta} = 0.5$) and measuring both the
H$\beta$ and Ly$\alpha$ luminosity functions in a 
$\sim 100,000$~Mpc$^3$ volume of space.  We show that in our redshift window, 
the volumetric Ly$\alpha$ escape fraction is at most
$4.4^{+2.1}_{-1.2}\%$, with an 
additional systematic $\sim 25\%$ uncertainty associated with our estimate of 
extinction.  Finally, we demonstrate that the bulk of the epoch's star-forming 
galaxies have Ly$\alpha$ emission line optical depths that are significantly 
greater than that for the underlying UV continuum.  In our predominantly 
[O~III] $\lambda 5007$-selected sample of galaxies, resonant scattering must 
be important for the escape of Ly$\alpha$ photons.

\end{abstract}


\section{Introduction}
\label{sec:Intro}
Ly$\alpha$ is the most common electronic transition in the universe.
Most often, it is a product of the photo-ionizing photons emitted by young
stars:  as recombining electrons cascade through the energy levels, they
are funneled into hydrogen's $n = 2$ state by the high optical depth of
the interstellar medium to Lyman series transitions.  The result is that 
strong Ly$\alpha$ is a signature of star formation, and indeed, 
\citet{partridge+67} noted that this feature may be our best probe for 
identifying galaxies in the act of formation.

Due to the resonant nature of the line, a typical Ly$\alpha$
photon must undergo tens or even hundreds of absorptions and re-emissions 
before escaping into intergalactic space.  Consequently, the radiative transfer 
of this line is quite complex, and even a small amount of dust can break the
chain of interactions which is necessary for its escape.  This fact
is reflected in the observed redshift evolution of Ly$\alpha$ emitting
galaxies (LAEs):  in the nearby universe, such objects are quite rare,
but between $z \sim 0.3$ \citep{deharveng+08, cowie+10} and
$z \sim 3$ there is a strong increase in both the number density 
of Ly$\alpha$ emitters and their characteristic luminosity
\citep{gronwall+07, ouchi+08, hayes+10, blanc+11, cassata+11, 
ciardullo+12, wold+14}.

Three-dimensional radiative transfer models have demonstrated that the
Ly$\alpha$ emission line can contain a great deal of information about 
the distribution of a galaxy's ISM, its surrounding 
circum-galactic medium, and the physics of its on-going star formation
\citep[e.g.,][]{neufeld91, hansen+06, verhamme+06, schaerer+11}. 
However, to extract this information, one needs accurate measurements of the 
fraction of Ly$\alpha$ photons escaping the galaxy (\fesc), and the profile 
of the emission line.

Over the past decade, there have been numerous attempts to estimate \fesc\  
in the normal (non-AGN) galaxies of the distant ($z \gtrsim 2$) universe, 
mostly by comparing Ly$\alpha$ to measurements of galactic emission in the 
rest-frame UV \citep[e.g.,][]{gronwall+07, ouchi+08, nilsson+09, blanc+11}, the 
far-infrared \citep{wardlow+14}, or the X-ray \citep{zheng+12}.  The premise 
behind these measurements is straightforward: like Ly$\alpha$, the strength
of a galaxy's UV, far-IR, and X-ray emission all depend in some way on the
existence of young stars and star formation.  Consequently, the ratio of 
Ly$\alpha$ to these quantities should yield a measure how efficiently 
Ly$\alpha$ is escaping its environment.  Using a compilation of such 
measurements, \citet{hayes+11} determined that, over time, the ``volumetric'' 
Ly$\alpha$ escape fraction of the universe has declined monotonically, from 
$\sim 40\%$ at $z \sim 6$ to $\sim 1\%$ locally.  This evolution 
is consistent with models in which Ly$\alpha$ is quenched by dust, which 
slowly builds up as the universe ages.

There is, however, one difficulty with this analysis: all the star-formation
rate tracers listed above are somewhat indirect and rely on empirical
calibrations derived from galaxies in the local $z \sim 0$ universe.  
For example, not only is the observed UV luminosity of a galaxy extremely 
sensitive to the effects of internal extinction, which may depend on such 
factors as star formation rate, inclination, and redshift 
\citep[e.g.,][]{buat+11, kriek+13}, but it also arises from a stellar 
population that is slightly different from that which is producing the 
ionizing photons.  Although both the UV continuum and Ly$\alpha$ are generated 
by the flux from hot, young stars, Ly$\alpha$ is excited by the far-UV 
emission of stars with $M \gtrsim 15 M_{\odot}$, whereas the UV light 
originates in the atmospheres of $M \gtrsim 5 M_{\odot}$ objects 
\citep[e.g.,][]{kennicutt+12}.   Thus, the ratio of the two quantities is 
subject to shifts in the initial mass function, metallicity, extinction law, 
and the timescale over which star formation is occurring.   Indeed, 
\citet{zeimann-1} has examined these effects and has shown that, even if both 
the star-formation rate (SFR) and extinction are well determined, measurements 
of the rest-frame UV in galaxies at $z \sim 2$ will underestimate the flux of 
ionizing photons by almost a factor of two.

To overcome the need for empirical calibrations, one requires a more direct
probe of the ionizing flux from hot, young stars.  Since Ly$\alpha$ is 
produced by transitions out of the $n = 2$ state of hydrogen, the
best possible tracer of its intrinsic strength is one which measures the
preceding transitions into the $n = 2$ state.  To date, only one such
investigation of this type has been made in the $z \gtrsim 2$
universe.  By performing narrow-band surveys for $z \sim 2.2$ galaxies in both
H$\alpha$ and Ly$\alpha$, \citet{hayes+10} was able to fix the epoch's
volumetric Ly$\alpha$ escape fraction at $5.3 \pm 3.8$\%.  
However, the precision of this measurement was limited by the 
survey's small volume ($\sim 5440$~Mpc$^3$), and the dearth of galaxies 
brighter than $L^*$.

To improve upon this situation, we have combined the data from 
four recent surveys: 3D-HST and AGHAST \citep{3DHST, weiner+14},
the Pilot Survey for HETDEX, the 
Hobby-Eberly Telescope Dark Energy Experiment \citep[HPS;][]{hetdex-1}, and 
the 3727~\AA\ narrow-band observations of the Chandra Deep Field South 
\citep[CDF-S;][]{guaita+10, ciardullo+12}.  The first two of these 
studies unambiguously measures total H$\beta$ fluxes in the redshift range 
$1.90 < z < 2.35$ via WFC3 grism observations with the {\sl Hubble Space 
Telescope;}  the latter two provide Ly$\alpha$ measurements 
(or upper limits) for many of these same galaxies via integral field 
spectroscopy and narrow-band imaging.  By comparing their data products, we 
can place constraints on \fesc\  via statistically complete samples of 
star-forming galaxies in the GOODS-N, GOODS-S, and COSMOS fields.  

In Section~2, we describe the observational data and detail the procedures 
used for identifying and measuring H$\beta$ and Ly$\alpha$ in our target
fields.  In Section~3, we use these data to measure (or place limits on)
the Ly$\alpha$/H$\beta$ ratio of 73 galaxies in the redshift range
$1.92 < z < 2.35$.  By converting stellar reddenings into nebular extinctions
via the empirical \citet{calzetti01} obscuration law, we show that the
typical Ly$\alpha$ escape fraction of these galaxies is just a few percent.  
In Section 4, we consider the epoch's volumetric Ly$\alpha$ escape fraction
by deriving the H$\beta$ and Ly$\alpha$ luminosity functions for a
$\sim 100,000$~Mpc$^{-3}$ volume of space.  After correcting for nebular
H$\beta$ extinction, we demonstrate that, at most, only 
$4.4^{+2.1}_{-1.2}\%$ of the Ly$\alpha$ photons escape their galaxies, and 
argue that any systematic error associated with this measurement must be less 
than $\sim 25\%$.  We conclude by discussing this measurement, and the
implications it has for the evolution of galaxies.  

For this paper, we assume a $\Lambda$CDM cosmology, with
$\Omega_{\Lambda} = 0.7$, $\Omega_M = 0.3$ and
$H_0 = 70$~km~s$^{-1}$~Mpc$^{-1}$ \citep{hinshaw+13}.

\section{The Sample}
\label{sec:Sample}
We begin our analysis with a sample of $z \sim 2$ galaxies observed with the
G141 near-IR grism of the {\sl Hubble Space Telescope's\/} Wide Field 
Camera 3 (GO programs 11600, 12177, and 12328).  This dataset, 
which is the product of the 3D-HST \citep{3DHST} and AGHAST 
\citep{weiner+14} surveys, 
consists of $R \sim 130$ slitless spectroscopy over the wavelength 
range $1.08~\mu{\rm m} < \lambda < 1.68~\mu$m, and records total emission line
fluxes over 625~arcmin$^2$ of sky.  Tens of thousands of spectra are observable
on these images, but of special interest are those produced by galaxies 
in the redshift range $1.90 < z < 2.35$, where the emission lines of 
[O~II] $\lambda 3727$, H$\beta$, and the distinctively-shaped [O~III] 
blended doublet $\lambda\lambda 4959,5007$ are simultaneously present in
the bandpass.  For these objects, redshift determinations are unambiguous, and 
total H$\beta$ fluxes can be measured to a 50\% completeness flux limit of 
$F \sim 10^{-17}$~ergs~cm$^{-2}$~s$^{-1}$, independent of redshift.  
\citet{zeimann-1} has used
these data to measure the H$\beta$ fluxes of 260 [O~II] and [O~III] selected 
galaxies in the GOODS-S, GOODS-N \citep{GOODS}, and COSMOS \citep{COSMOS} 
fields, while \citet{gebhardt+14} has derived metallicities and masses for 
these systems.   
Comparisons with the deep X-ray surveys of the regions
\citep{elvis+09, alexander+03, xue+11} confirm that the vast majority 
of these objects are normal galaxies with 
star-formation rates between $\sim 1$ and $\sim 200 \, M_{\odot}$~yr$^{-1}$ and
no evidence of AGN activity \citep{zeimann-1}.  Any H$\beta$ source projected
within $2\farcs 5$ of a cataloged X-ray position has been excluded from 
our analysis.

Our Ly$\alpha$ measurements (and upper limits) come principally from HPS,
a blind integral-field spectroscopic study of four areas of sky, including 
COSMOS and GOODS-N\null.  A full description of this survey and its data 
products is given by \citet{hetdex-1}, but in brief, a square 246-fiber array 
mounted on the Harlan J. Smith 2.7-m telescope at McDonald Observatory was 
coupled to the $R \sim 850$ George and Cynthia Mitchell Spectrograph, a 
proto-type of the Visible Integral-field Replicable Unit Spectrograph 
(VIRUS-P) designed for HETDEX \citep{VIRUS}.  At the focal plane, each fiber 
subtends an angle $4\farcs 2$ in diameter, enabling simultaneous spectroscopy 
of $\sim 2.7$~arcmin$^2$ of sky between the wavelengths 3550~\AA\ and 
5800~\AA\null.  At $z \sim 2.2$, 50\% of the survey's pointings 
reach a $5\, \sigma$ monochromatic flux limit of 
$1.3 \times 10^{-16}$~ergs~cm$^{-2}$~s$^{-1}$ (or $\log L({\rm Ly}\alpha) = 
42.68$~ergs~cm$^{-2}$~s$^{-1}$)  and 90\% reach 
$2.5 \times 10^{-16}$~ergs~cm$^{-2}$~s$^{-1}$
($\log L({\rm Ly}\alpha) = 42.96$~ergs~cm$^{-2}$~s$^{-1}$).
Above these flux limits,
the HPS's recovery fraction of emission lines is greater
than 95\% for emission line equivalent widths greater than 5~\AA, and
better than 90\% for equivalent widths as small as 1~\AA\ \citep{hetdex-1}.
Moreover, because the VIRUS-P spectrograph's response increases
rapidly towards the red, the luminosity limits of this survey are roughly
constant throughout our redshift range of our observations
\citep{blanc+11}. 

In total, the HPS survey covered 169~arcmin$^2$, with 107~arcmin$^2$
in the COSMOS and GOODS-N regions.  Roughly 76~arcmin$^2$ of this 
area overlaps the fields studied by {\sl HST,} 
with $\sim 85\%$ of the overlap region useable for science 
(see Section~\ref{subsec:Hbeta_LF}).
Thus, the intersection of two surveys encompasses a co-moving volume of 
$\sim 93,000$~Mpc$^3$ between $1.916 < z < 2.350$.

To compare the HPS Ly$\alpha$ measurements with the H$\beta$ fluxes from the
{\sl HST\/} grism, we began by examining the VIRUS-P spectra at the location 
of every galaxy in the {\sl HST\/} emission-line selected sample.  Since the 
original HPS survey ignored all detections below $5 \, \sigma$ confidence, we 
re-measured these spectra, mimicking the procedures of \citet{hetdex-1} using
a $6\arcsec$ radius aperture centered on the position of each H$\beta$ source.
This spectrophotometry did present some challenges.  Because the dither 
pattern of the HPS placed roughly two of the $4.2\arcsec$ diameter fibers at 
any given position, we had to precisely compute the fraction of each fiber 
falling within any given aperture.  This calculation was done by summing the 
total fiber area within the designated aperture and then normalizing the flux 
by this aperture area.  In addition, although we did have prior knowledge 
of the approximate redshift of each source, the limited spectral resolution
of the {\sl HST\/} grism
\citep[$\Delta z \sim 0.005$ at $z \sim 2$;][]{3DHST, colbert+13}
prevented us 
from knowing the exact wavelength of the corresponding Ly$\alpha$ line.  We 
therefore searched a spectral window about the H$\beta$-defined wavelength
of Ly$\alpha$ that was six times the wavelength resolution of the grism,
($\pm 6 \cdot 1216~{\rm\AA\ } \cdot (1 + z) \cdot \Delta z \sim 109$~\AA),
summing up the putative Ly$\alpha$ flux in a series of 4.2~\AA\  bandpasses 
(i.e., twice the spectral resolution of the instrument).   The total flux
and noise were then corrected for flux losses due to the fixed spatial
and spectral aperture by assuming an effective PSF through the $4\farcs 2$
fibers of $6\arcsec$ FWHM and adopting an 8~\AA\ FWHM for the line profile.
Finally, we searched for sources with a signal-to-noise detection greater
than 3 within the spectral bandpass.  This procedure recovered the fluxes
of all the $5 \, \sigma$ Ly$\alpha$ detections found by \citet{hetdex-1} in 
our H$\beta$ redshift window, and identified one additional H$\beta$ 
counterpart with a Ly$\alpha$ $S/N$ of $\sim 4$.  For the remaining H$\beta$ 
sources we used the $3 \, \sigma$ Ly$\alpha$ limits in our analysis.

To supplement the HPS data, we also used a second source of Ly$\alpha$
measurements:  the deep, narrow-band observations of the Extended Chandra Deep 
Field South.  \citet{guaita+10} imaged this field with the CTIO 4-m
and Mosaic camera for 28.17 hours through a 50~\AA\ wide interference 
filter centered at 3727~\AA, and obtained a sample of over
200~Ly$\alpha$ emitting objects in the redshift range $2.04 \lesssim z \lesssim
2.08$.  [See \citet{ciardullo+12} for more details on this dataset.]
A subset of these sources fall in the GOODS-S region surveyed by
3D-HST, thus allowing us to increase the number of Ly$\alpha$ emitters
with reliable H$\beta$ constraints.  Because these narrow-band observations 
sample only a small redshift slice of the universe, the actual volume 
covered by the GOODS-S data is rather small, just $\sim 15,000$~Mpc$^3$, or 
$\sim 15\%$ that of the overlap region between HPS and the fields surveyed
by the {\sl HST\/} grism.  Also, due to the redshift uncertainty 
associated with the {\sl HST\/} grism measurements, the Ly$\alpha$ emission
from any given H$\beta$ source with $2.02 \lesssim z_{\rm grism} 
\lesssim 2.09$ may, or may not fall within the narrow-band filter's bandpass.
This ambiguity, plus the flux errors introduced by the filter's
Gaussian-shaped transmission curve, complicates the interpretation of 
these photometric measurements.  Nevertheless, the narrow-band Ly$\alpha$ 
data do reach significantly deeper than the HPS spectra (to a 
90\% completeness limit of $2 \times 10^{-17}$~ergs~cm$^{-2}$~s$^{-1}$) and 
are useful, for both probing the faint end of the LAE luminosity function,
and performing the inverse experiment of measuring \fesc\ for targets 
where Ly$\alpha$ is already detected.

Finally, to complete our sample, we performed a reverse search and examined
the {\sl HST\/} grism frames for evidence of H$\beta$ emission at the position 
of known Ly$\alpha$ sources.  For several objects, this proved to be impossible
due to contamination from overlapping spectra. However, in a few cases, 
we were able to extract the H$\beta$ fluxes for the LAE candidate, thereby 
reconfirming its existence.  Our final sample in the area of HPS/HST 
overlap therefore consists of 54 H$\beta$ emitting galaxies 
and 13 Ly$\alpha$ emitters, with four objects detected in both surveys.
The GOODS-S region contains 24 galaxies with grism-based redshifts between 
$2.03 \leq z \leq 2.08$, though only 13 have redshifts that fall within
the effective volume surveyed by the Gaussian-shaped narrow-band filter 
\citep[defined by the filter's full-width at two-thirds maximum; 
see][for the details of this calculation]{gronwall+07}.  Of the 10 narrow-band 
selected Ly$\alpha$ emitters falling within the 3D-HST survey area, 8 have 
H$\beta$ detections.  A list of these objects appears in 
Table~\ref{tab:galaxies}. 

\section{Individual Ly$\alpha$/H$\beta$ Ratios}
\label{sec:Comparison}
\begin{figure*}[t]
\figurenum{1}
\epsscale{0.8}
\plotone{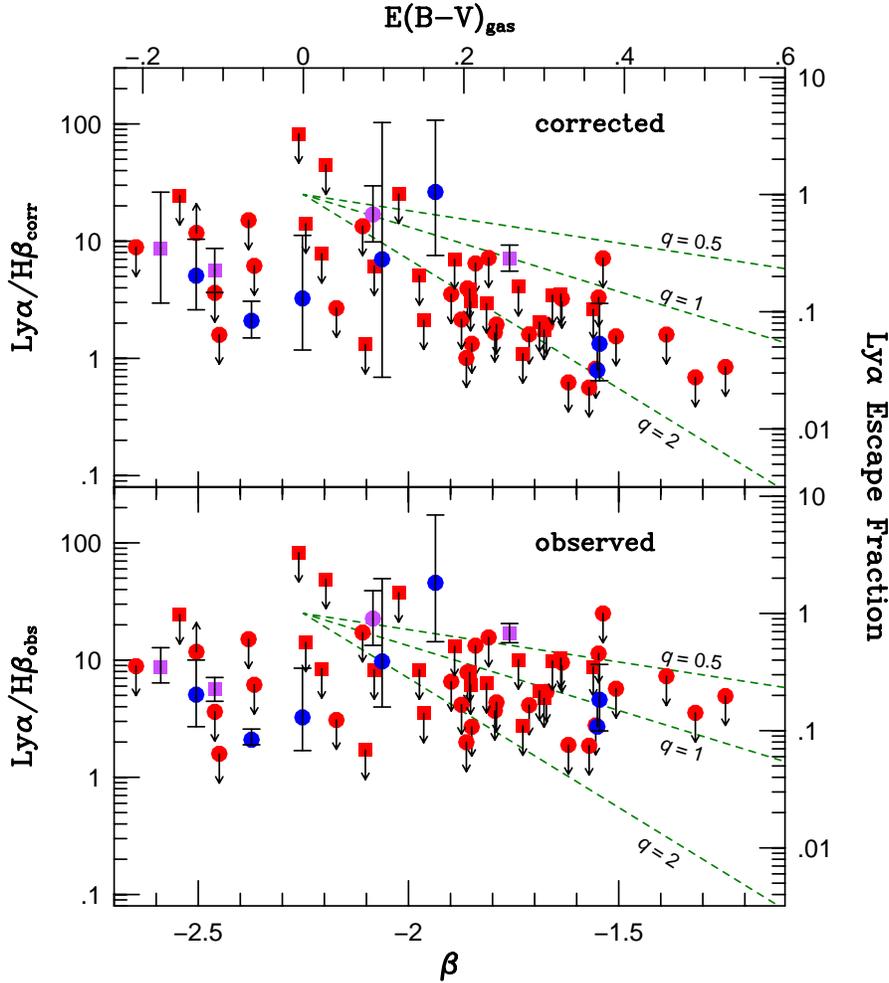}
\caption{Ly$\alpha$/H$\beta$ ratios for $1.92 < z < 2.35$ {\sl HST-}grism 
galaxies plotted against the power-law slope of the galaxies' rest-frame UV 
continua ($\beta$) and their corresponding nebular extinction [$E(B-V)$], 
assuming a \citet{calzetti01} obscuration relation.  The red points show 
Ly$\alpha$ upper limits from the integral-field spectroscopy of the HETDEX 
Pilot Survey, with the circles representing COSMOS field objects and the 
squares denoting sources in GOODS-N\null.  The purple points indicate HPS 
objects with both Ly$\alpha$ and H$\beta$ detections.  For comparison, the 
blue points are the ratios of GOODS-S LAEs detected via narrow-band imaging.  
The bottom panel presents the observed data, while the top panel corrects the 
H$\beta$ line fluxes for extinction using the slope of the UV continuum and 
the relation between stellar reddening and obscuration given by 
\citet{calzetti01}.  (Objects with $\beta < -2.25$ have been assigned zero 
reddening.)  The green dashed lines show where the optical depth to Ly$\alpha$ 
is 0.5, 1.0, and 2.0 times that of the continuum at 1216~\AA\null.  After 
correcting for extinction, most, if not all the galaxies lie below this line, 
demonstrating the importance of resonant Ly$\alpha$ scattering within these 
galaxies.}
\label{fig:ratio}
\end{figure*}

The lower panel of Figure~\ref{fig:ratio} compares our {\sl HST\/} H$\beta$ 
fluxes to Ly$\alpha$ measurements and limits made via HPS spectroscopy.
The measurements are plotted against $\beta$, the slope of the rest-frame
UV continuum between 1250~\AA\ and 2600~\AA, 
as derived from the deep multicolor photometry compiled and
homogenized by \citet{skelton+14}.  Also plotted are the LAEs found 
in the GOODS-S field.   From the figure, it is immediately obvious that for 
most galaxies, Ly$\alpha$ is below the detection limit of the ground-based 
surveys.  As stated above, only four of the 54 H$\beta$ emitting galaxies were 
detected with the HPS spectroscopy; this agrees with the results of 
\citet{hayes+10}, who found little overlap in the blind samples of Ly$\alpha$ 
and H$\alpha$ emitters at $z = 2.2$.  The fraction of Ly$\alpha$ recoveries is 
greater for the deeper GOODS-S data, but since it is not known exactly where
these objects fall on the narrow-band filter's transmission curve, the
uncertainties associated with their Ly$\alpha$ fluxes are generally large.
Nevertheless, these objects do suggest the existence of an upper limit to
the Ly$\alpha$/H$\beta$ ratios of galaxies.

Of course, the values shown in the lower panel of Figure~\ref{fig:ratio} do 
not reflect the true ratio of these emission lines.  Before we can use these 
limits to infer Ly$\alpha$ escape fractions, we must correct H$\beta$ for the
effects of internal extinction.  For the dataset under consideration, this
is not straightforward.  The {\sl HST\/} grism spectra do not extend to 
$z \sim 2$ H$\alpha$, and H$\gamma$ is generally too faint (and too close in 
wavelength to H$\beta$) to constrain the Balmer decrement.  Consequently, we 
have no direct measure of the extinction affecting the objects' recombination 
lines.  We do, however, have access to a measure of {\it stellar\/} reddening, 
as each field has deep multicolor photometry that extends throughout the 
rest-frame UV, from 1250~\AA\ to 2600~\AA\ \citep{skelton+14}.  
\citet{calzetti01} has shown that over this spectral range, the intrinsic 
slope of the stellar continuum of star-forming galaxies is very nearly a power
law, with $F_{\lambda} \propto \lambda^{\beta_0}$.  For steady-state
star-formation extending over $\sim 0.5$~Gyr, $\beta_0 \approx -2.25$, 
while for extremely young starbursts, $\beta_0$ may be as steep as $-2.70$.  
If the rest-frame UV slope of a star-forming galaxy is observed to be flatter 
than this value, the most likely explanation is reddening due to dust.

We next need to know how the observed slope of the UV continuum 
translates into nebular extinction.  As has been noted many times in the 
literature, the distribution of stars within a galaxy is generally wider than
that of the dust, and the latter is often associated with individual H~II
regions \citep[e.g.,][]{cha-fall00}.  As a result, emission-line gas is
usually extinguished more than the stars.  By observing 8 starburst galaxies 
in the local universe, \citet{calzetti01} concluded that $E(B-V)_{\rm stars} 
= 0.44 E(B-V)_{\rm gas}$ and that the slope of the
rest-frame UV continuum is related to the total stellar extinction (in
magnitudes) at 1600~\AA\  and the total nebular H$\beta$ extinction by
\begin{equation}
A_{1600} = \kappa_{\beta} \, \Delta \beta \ \ \ {\rm and} \ \ \ 
A_{{\rm H}\beta} = \zeta_{{\rm H}\beta} \, A_{1600} 
\label{eq:calzetti}
\end{equation}
with $\kappa_{\beta} = 2.31$ and $\zeta_{{\rm H}\beta} = 0.83$.
Without direct measurements of the Balmer decrement for large samples of
$z \sim 2$ galaxies, it is impossible to confirm this relation for the
objects in our sample.  Nevertheless, using the same G141 grism data 
studied here, \citet{zeimann-1} showed that, indeed, a power-law fit is good
representation of the stellar continuum, and the product
\begin{equation}
0.4 \times \left( 1 - \zeta_{{\rm H}\beta} \right) \kappa_{\beta} =
0.155 \pm 0.043
\label{eq:zeimann}
\end{equation}
is at least consistent with the value of 0.162 expected from the 
\citet{calzetti01} obscuration relation.  This law is also supported by several
recent surveys \citep[e.g.,][]{forster+09, mannucci+09, holden+14}, though 
others have found $E(B-V)_{\rm gas}>E(B-V)_{\rm stars}>0.44 E(B-V)_{\rm gas}$ 
\citep{wuyts+13, price+14}.  Nevertheless, it is reasonable to conclude
that a \citet{calzetti01} relation, at least in the statistical sense, is 
applicable to the galaxies in our sample.  

The top panel of Figure~\ref{fig:ratio} repeats our Ly$\alpha$/H$\beta$
comparison, but with H$\beta$ corrected for extinction via the 
\citet{calzetti01} obscuration relation and the assumption that 
$\beta_0 = -2.25$.  These values can easily be translated into escape 
fractions.  Under Case~B recombination, every ionization results in the 
creation of a Balmer-line photon, with $\sim 11.5\%$ of these photons coming 
via an $n = 4$ to $n = 2$ transision 
\citep[i.e., H$\beta$;][]{pengelly64, AGN3}.  
Three quarters of these Balmer transitions land in 
a $2 P$ orbital and immediately decay to the ground state via the emission of 
Ly$\alpha$; the other 25\% of the electrons become temporarily trapped in the 
$2 S$ state before decaying to $n = 1$ via two-photon emission.  Thus, 
under normal conditions, the Ly$\alpha$/H$\beta$ ratio should be
\begin{equation}
R = \frac{I({\rm Ly}\alpha)}{I({\rm H}\beta)} = \frac{3}{4} \,
\frac{\alpha_B}{\alpha_{{\rm H}\beta}^{\rm eff}} \,
\frac{h \nu_{{\rm Ly}\alpha}}{h \nu_{{\rm H}\beta}} \approx 25
\label{eq:ratio}
\end{equation}
where $\alpha_B$ is the Case~B recombination coefficient ($2.59 \times 
10^{-13}$~cm$^3$~s$^{-1}$ at 10,000~K) and 
$\alpha_{{\rm H}\beta}^{\rm eff}$ is the effective Case B recombination 
coefficient for H$\beta$ \citep[$3.03 \times 10^{-14}$~cm$^3$~s$^{-1}$ at
$T = 10000$~K;][]{pengelly64, AGN3}. 

The escape fractions shown in Figure~\ref{fig:ratio} used $R = 25$ as the 
intrinsic ratio of Ly$\alpha$ to H$\beta$. In practice, however, these escape 
fractions are upper limits.  If the Case~B condition is relaxed so that the 
Lyman continuum is optically thin, then the Ly$\alpha$/H$\beta$ ratio may be 
boosted to values as large as $R \sim 33$, thereby lowering \fesc.  Most
evidence suggests that at the redshifts considered here, the escape fraction 
of Lyman continuum photons is at most a couple of percent
\citep[e.g.,][]{chen+07, iwata+09, vanzella+10}, but this possibility cannot 
be excluded.  Similarly, if the ISM density approaches 
$n_e / \sqrt{T_e/10^4} \gtrsim 10^4$~cm$^{-3}$ collisions will redistribute 
$2 S$ electrons into the $2 P$ state, again enhancing Ly$\alpha$ 
relative to H$\beta$.  In the galaxies of the local universe, most H~II 
regions have electron densities between 1 and 100~cm$^{-3}$ 
\citep[e.g.,][]{gutierrez+10}, though this number may
be slightly larger in dwarf systems \citep[e.g.,][]{hunter+99}.  Shocks
or very high densities (such as in the broad-line regions of AGN) may also
increase Ly$\alpha$ relative to H$\beta$ by creating an environment where
the $n=2$ state of neutral material is collisionally populated.  
Finally, our estimates of nebular reddening assume that the observed
$z \sim 2$ galaxies have been undergoing vigorous star formation for at least 
$\sim 0.5$~Gyr, so that $\beta_0 = -2.25$.  Since all our grism-selected 
objects have very high H$\beta$ equivalent widths, the intrinsic slopes of 
their rest-frame UV continua are unlikely to be flatter than this value
\citep{zeimann-1}.  However, $\beta_0$ could be steeper, up to $-2.7$ in the
extreme, if the systems have only recently begun their star forming activity
\citep{calzetti01}.  In this case, our reddening estimates would be
underestimated, H$\beta$ would be enhanced, and, once again, our inferred 
ratios of Ly$\alpha$ to H$\beta$ would need to be reduced.   

In summary, our adopted values of $R = 25$ and $\beta_0 = -2.25$ both produce 
upper limits for the Ly$\alpha$ escape fraction.  Any deviations from 
simple Case B recombination or steady-state star formation will only serve
to reduce \fesc.   While factor of $\sim 2$ errors are theoretically
possible for systems younger than $\sim 2$~Myr, in most cases, any systematic
error associated with our escape fraction measurements should be less than 
$\sim 30\%$.

As the upper panel of Figure~\ref{fig:ratio} illustrates, the median H$\beta$ 
emitter detected by the {\sl HST\/} grism has an escape fraction below 7\%, 
with no statistical difference between the results of the two HPS fields.
Note that our constraints on \fesc\  become progressively stronger as the
slope of the UV continuum becomes redder.  This is simply the result of our
extinction law:  as the stellar reddening increases, so does the assumed
extinction correction for H$\beta$.  The implied increase in H$\beta$ then 
translates into a decreased limit for the Ly$\alpha$/H$\beta$ ratio.

Another way to view the data of Figure~\ref{fig:ratio} is through the
mathematics of survival analysis \citep{cohen91, lee-wang03, feigelson-babu}.  
Several authors have shown that the distribution of Ly$\alpha$ equivalent
widths is exponential in form above a rest-frame equivalent width of 
$\sim 20$~\AA\ \citep{gronwall+07, nilsson+09, ciardullo+12}.  If the
same is true for Ly$\alpha$ escape fractions, then the computation of the
median escape fraction is straightforward.   The result is a most likely 
median value of $f_{\rm esc}^{{\rm Ly}\alpha} = 
5.9^{+1.0}_{-0.9}$\%, where the errors are
computed via a Markov Chain Monte Carlo sampler \citep{emcee}.  Of course,
since this calculation depends on the underlying shape of the distribution, 
the formal errors on \fesc\  underestimate the true uncertainty 
in the measurement.

Perhaps the most instructive way to interpret Figure~\ref{fig:ratio} is
through the geometric parameter $q$, which has been defined by 
\citet{finkelstein+08} as the ratio of the optical depth of Ly$\alpha$ 
to that of the stellar continuum at 1216~\AA\null.  In objects where
resonant scattering is important and the escape path for Ly$\alpha$
photons is long, the likelihood of such a photon encountering a dust grain
is large and $q > 1$.  Conversely, if $q \sim 1$, Ly$\alpha$ must
have a relatively direct escape route from the galaxy, with a path-length 
similar to that of the star light.  Values of $q < 1$ point to either
anisotropic emission, a spatial offset between the points of origin for
Ly$\alpha$ and the continuum, a clumpy ISM \citep{neufeld91, hansen+06}, or
significant deviations from Case~B recombination.

The dashed lines Figure~\ref{fig:ratio} display the $q = 0.5, 1$ and 2
relations under the assumption that a \citet{calzetti01} obscuration relation 
holds, with $A_{1216} = 2.77 \, \Delta \beta$.  As expected, most of the 
sources have upper limits that are larger than $q = 1$, demonstrating that
resonant scattering of Ly$\alpha$ photons within these galaxies is 
significant.  Perhaps as importantly, none of the objects show values of 
$q$ significantly less than one.  Several studies have found that luminous
Ly$\alpha$ emitting galaxies tend to cluster about the $q = 1$ line
\citep[e.g.,][]{finkelstein+09, finkelstein+11, blanc+11, nakajima+12}, and
\citet{hagen+14a} found that small values of $q$ are common in
compact, low-mass LAEs.  But these analyses were for systems selected
via their strong Ly$\alpha$ emission.  The sources analyzed here
were primarily chosen via their bright [O~III] (or 
[O~II]) lines.  In these more normal star-forming galaxies,
$q$ is generally greater than one, demonstrating that Ly$\alpha$ has a 
difficult time escaping its immediate environment.

\section{The Global Escape Fraction}
\label{sec:LF}
An alternative approach to exploring the escape fraction of Ly$\alpha$ photons 
from the $z \sim 2$ universe is to do so globally, via a comparison of the 
H$\beta$ and Ly$\alpha$ luminosity functions.  The $\sim 100,000$~Mpc$^3$ 
volume of space surveyed for both H$\beta$ and Ly$\alpha$ contains 
67 galaxies detected in H$\beta$ and 23 identifiable via 
Ly$\alpha$.  These numbers are more than sufficient for defining the 
emission-line luminosity functions of the two galaxy populations and 
integrating for their total luminosity density.

\subsection{The H$\beta$ Luminosity Function}
\label{subsec:Hbeta_LF}
In order to calculate the $z \sim 2$ H$\beta$ luminosity function, we must first
estimate the effective area of overlap between the HPS and the {\sl HST\/} 
grism observations.  This is not just a simple geometry problem, since, as 
with all slitless spectroscopy,  overlapping sources render a fraction of the 
survey area unusable.   \citet{zeimann-1} estimated this fraction via a series 
of Monte Carlo experiments, in which realistic magnitude and positional 
distributions were used to create simulated frames, which were then 
``observed'' in the same manner as the original data.   At each point on the 
simulated frames, the amount of contamination was compared to the local sky 
noise, and all regions where the systematics of spectral subtraction 
were greater than this noise were excluded from consideration.  Since sky noise 
is the dominant source of uncertainty for all $z \sim 2$ observations, this 
procedure generated a reliable statistical measure of the grism survey's 
effective area.  Based on the results of these simulations, we reduced the 
geometric HPS/HST overlap region by 15\% to 65~arcmin$^2$, and used this 
new area in our calculation of survey volume.

\begin{figure*}[t]
\figurenum{2}
\epsscale{1.0}
\plotone{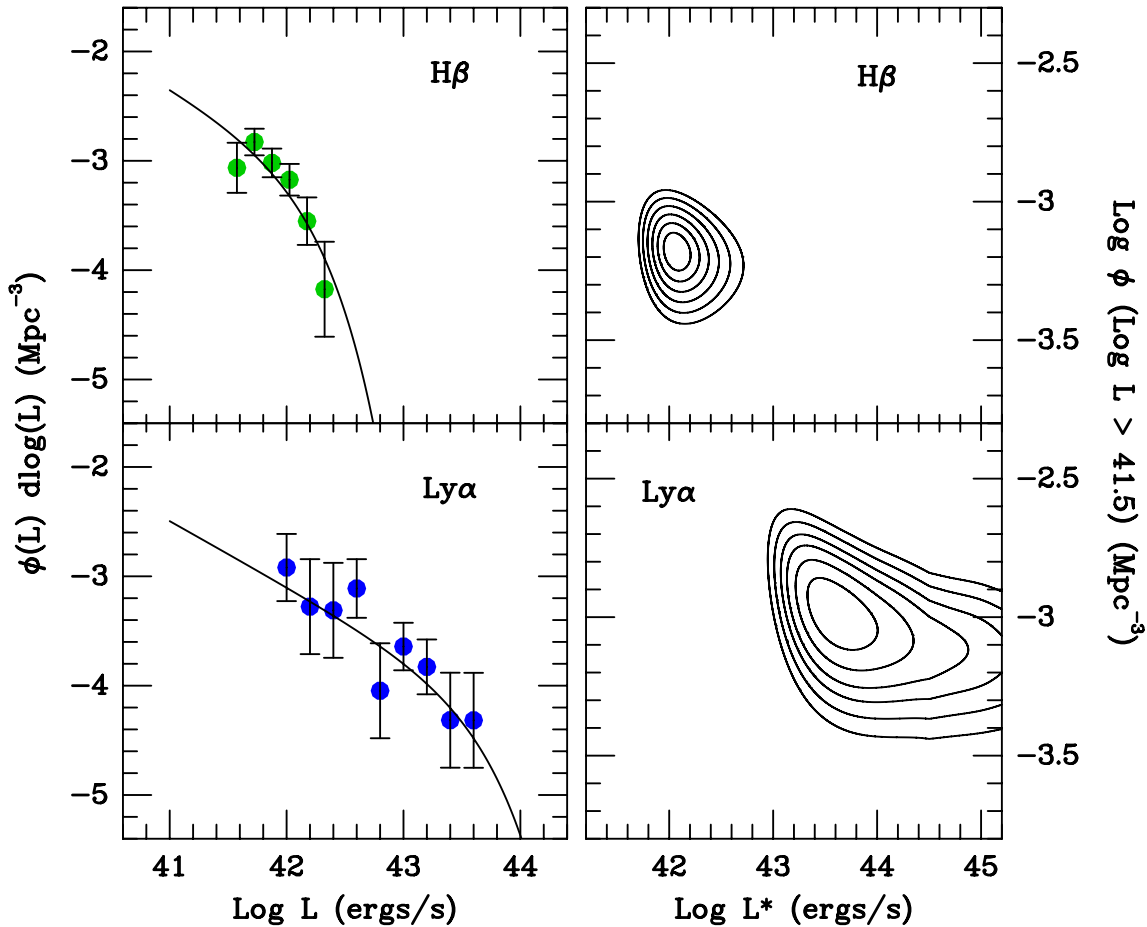}
\caption{The H$\beta$ and Ly$\alpha$ luminosity functions in the 
$1.92 < z < 2.35$ regions surveyed for both emission lines.  The left-hand 
panels show the observed luminosity functions (binned into 0.15~dex and 
0.20~dex intervals for H$\beta$ and Ly$\alpha$, respectively), and the 
best-fit \citet{schechter} functions.  The right panels display the likelihood 
contours (in $0.5 \, \sigma$ intervals) of the fitted functions, with 
$\log L^*$ on the $x$-axis and the number density of galaxies brighter than 
$\log L = 41.5$ (ergs~s$^{-1}$) on the $y$-axis.  The slope of the faint end 
of the luminosity function has been fixed at $\alpha = -1.6$.}
\label{fig:lf}
\end{figure*}

To measure the total H$\beta$ luminosity density at $z \sim 2$, we began
by following the procedures described by \citet{blanc+11} and applied
the $1/V_{\rm max}$ technique \citep{schmidt68, huchra73} to the H$\beta$
sources found in COSMOS, GOODS-N, and GOODS-S\null.  As described in
\citet{zeimann-1}, the completeness fraction of our {\sl HST\/} grism sample, 
as a function of H$\beta$ flux, has been calculated via a series of Monte
Carlo simulations; in summary, the 50\% completeness limit for the
two GOODS fields (in ergs~cm$^{-2}$~s$^{-1}$) occurs at 
$\log F_{{\rm H}\beta} = -17.06$, while for
COSMOS, this limit is $\log F_{{\rm H}\beta} = -16.88$.  In the regions of
overlap between the H$\beta$ and Ly$\alpha$ surveys, 42 galaxies are present
above these completeness limits.   Using these data, 
we computed $V_{\rm max}$, the co-moving volume over which an object with 
H$\beta$ luminosity $L$ would be detected more than 50\% of the time.  The 
number density of galaxies in any absolute luminosity bin of
width $\Delta \log L$ is then
\begin{equation}
\phi(\log L) = \frac{1}{\Delta (\log L)} \, \eta \,
\sum_i \left\{ \frac{1}{V_{\rm max}(i)} \right\}
\label{eq:vmax}
\end{equation}
where $\eta$ is the inverse of the completeness function, and
the summation is performed over all galaxies with luminosities falling
within the bin.  The top-left panel of Figure~\ref{fig:lf} displays this 
function, where the uncertainties on the points are from Poissonian statistics 
only.

We next fit this function using the maximum-likelihood procedure detailed
in \citet{ciardullo+13}.  We began with the assumption that over the 
redshift range $1.92 < z < 2.35$, the H$\beta$ luminosity function 
can be modeled via a \citet{schechter} law
\begin{equation}
\phi(L/L^*) d(L/L^*) = \phi^* \left(L/L^*\right)^\alpha e^{-L/L^*} d(L/L^*)
\label{eq:schechter}
\end{equation}
with $L^*$ being the characteristic monochromatic luminosity of the
epoch.   The observed function, of course, does not follow this relation,
as incompleteness takes an ever increasing toll at fainter fluxes.  Hence we 
define $\phi^\prime(L,z)$ as the Schechter function modified by the 
flux-dependent completeness fraction at each redshift
as given by \citet{zeimann-1}.  
From Poissonian statistics, the probability of observing $n$ galaxies in any 
given luminosity and volume interval $\Delta L \, \Delta V$ is then 
\begin{equation}
p(n|\lambda) = \frac{\lambda^n e^{-\lambda}}{n!}
\label{eq:poisson}
\end{equation}
where the expectation value $\lambda = \phi^\prime(L,z) \Delta L \, \Delta V$.  
If we let these intervals become differentials, then the likelihood of drawing
an observed set of $N$ H$\beta$ luminosities from a given Schechter function
with parameters $L^*$, $\phi^*$, and $\alpha$ becomes
\begin{equation}
\ln P = -\int_{z_1}^{z_2} \int_{L_{\rm min}(z)}^\infty \phi^\prime(L,z)
dL \, dV + \sum_i^N \ln \phi^\prime(L_i,z_i)
\label{eq:likelihood}
\end{equation}
where, for purposes of our analysis, we define the lower limits of the 
luminosity integral by where the completeness fraction drops to 50\%.  
The top right panel of Figure~\ref{fig:lf} 
displays the likelihood contours in $L^*$ and $\phi^*$, with the 
faint-end slope fixed at $\alpha = -1.6$ for consistency with other studies
\citep{hayes+10, ly+11, sobral+13}.   To avoid the well-known
degeneracy between $L^*$ and $\phi^*$, the ordinate of the plot gives the
integral of the Schechter function (down to $\log L = 41.5$~ergs~s$^{-1}$), 
rather than the traditional coefficient, $\phi^*$.   The most likely 
solution, with $\log L^* = 42.07$ and $\phi_{\rm tot}(\log L > 41.5) 
= -3.18$~Mpc$^{-3}$, is displayed in the top left of Figure~\ref{fig:lf}.  For 
comparison, the most likely H$\beta$ luminosity function for the full 
sample of $1.90 < z < 2.35$ {\sl HST\/-}grism selected galaxies in COSMOS, 
GOODS-N, and GOODS-S has $\log L = 42.07$ and $\phi_{\rm tot}(\log L > 41.5) 
= -3.28$.

Table~\ref{tab:lf} lists the most-likely \citet{schechter} function 
parameters, their uncertainties, and the total integrated H$\beta$ luminosity 
density
\begin{equation}
\rho_{{\rm H}\beta} = \phi^* L^* \, \Gamma(\alpha+2) 
\label{eq:sch_int}
\end{equation}
for the $1.92 < z < 2.35$ regions of overlap between the {\sl HST\/} surveys, 
HPS, and the CDF-S narrow-band survey field.  Although this latter quantity 
does require the extrapolation of the \citet{schechter} function to zero
luminosity, this extension is not an important issue, as galaxies below our 
detection threshold are predicted to contribute only $\sim 20\%$ to the 
universe's total H$\beta$ luminosity density.  For reference, 
Table~\ref{tab:lf} also lists the H$\beta$ luminosity function parameters for 
the entire $1.90 < z < 2.35$ grism survey region of COSMOS and GOODS\null.  
Although this full volume is roughly four times larger than that for the just 
the regions of survey overlap, the luminosity function defined by its galaxies
is virtually identical to that of the smaller region.

The likelihoods given in Table~\ref{tab:lf} and displayed in 
Figure~\ref{fig:lf} represent only the formal statistical error of our fit. 
Not included are systematic uncertainties associated with the data themselves, 
most notably, with the completeness corrections.  Our estimates for the
survey area and completeness fraction are based on the simulations performed by 
\citet{zeimann-1}, who also confirmed that metallicity, equivalent width, and 
redshift are not important factors for determining the detectability of 
H$\beta$.  This analysis did not, however, take galaxy size into account,
and this can be an important factor for slitless spectroscopic surveys.  
Fortunately, for the dataset considered here, the effect is minor.  As
\citet{vanderwel+14} have shown, $\sim 84\%$ of massive ($M_* \gtrsim 10^{10} 
M_{\odot}$) late-type galaxies at $z = 2.25$ have effective radii smaller 
than $0 \farcs 45$, while all the H$\beta$ galaxies analyzed here have
$r_e < 0 \farcs 50$ \citep{hagen+14b}.  
In this regime, the completeness of an {\sl HST\/} grism survey is a very weak 
function of size \citep[see the analysis of][]{colbert+13}, and the simulations
of \citet{zeimann-1} should be valid.  Based on this result, our calculation 
of survey area and incompleteness fraction likely carries an additional error 
of $\sim 8\%$.

Similarly, because the {\sl HST\/} survey fields are relatively small, the 
effects of cosmic variance on our luminosity functions are non-negligible.  We 
can estimate this number using the cosmic variance calculator developed by 
\citet{trenti+08}, which combines Press-Schechter theory with N-body 
cosmological simulations to predict field-to-field fluctuations in arbitrary 
slices (or pencil beams) of the universe.  According to this estimator, our 
number counts in the three {\sl HST\/} fields carry an additional uncertainty 
of $\sim 12\%$, while that for just the 65~arcmin$^2$ region of overlap between
the {\sl HST-}grism surveys and our two sets of ground-based Ly$\alpha$ 
observations is $\sim 20\%$.  Of course, for estimates of the Ly$\alpha$ 
escape fraction, it is the field-to-field variations of galactic properties 
such as star formation rate and dust content which are the important parameters
for the calculation, not simply the number of galaxies present in the region.  
While we are less able to quantify this number, it seems likely that the 
effect is significantly less than $\sim 20\%$. 

The last piece of information needed to compute the H$\beta$ luminosity
density of the $z \sim 2$ universe is an estimate of H$\beta$ attenuation.
Once again, we are limited by the lack of direct knowledge of
the galaxies' nebular extinction.  
Measurements of the Balmer decrement
in the $z \sim 2$ universe are rare, especially for samples of 
emission-line selected galaxies.  (The closest comparison sample to
our own -- that created by \citet{dominguez+13} from 128 WFC3 grism-selected
galaxies in the $0.75 < z < 1.5$ universe --- has very large uncertainties, 
with a mean value of $c_{{\rm H}\beta} \sim 1.0 \pm 1.0$~dex.) 
Consequently, we must again compute the loss of H$\beta$ statistically
from measurements of the stellar continuum.  We did this by translating 
the UV slopes of each H$\beta$ source in our sample into a nebular extinction
using the obscuration relations of \citet{calzetti01}, and then examining the 
distribution of these extinctions.  If we just consider the sample of 42 
galaxies in the region of space surveyed for H$\beta$ and Ly$\alpha$, then the 
median logarithmic extinction at H$\beta$ is $c_{{\rm H}\beta} = 0.35$, while
the effective extinction, defined via the total amount of H$\beta$ luminosity
lost due to dust in all the galaxies, is $c_{{\rm H}\beta} = 0.44$.  
For comparison, the logarithmic extinctions found for all $1.90 < z <
2.35$ galaxies brighter than the 50\% H$\beta$ completeness limit in the 
{\sl HST\/} fields of COSMOS, GOODS-N, and GOODS-S is $c_{{\rm H}\beta} = 
0.31$ (median) and $c_{{\rm H}\beta} = 0.59$ (effective).  This difference
between the median and effective attenuation is not unexpected:  as
noted by many authors, reddening is strongly correlated with stellar mass and
therefore star-formation rate \citep[e.g.,][]{moustakas+06, garn-best, an+14}. 
Since the brightest H$\beta$ emitters are also those most heavily 
extinguished by dust, the effective extinction of the sample should be
larger than the median extinction.  For the remainder of this paper, we
will adopt $c_{{\rm H}\beta} = 0.5$ as the total logarithmic extinction at
H$\beta$, while noting that the uncertainty on this number is likely to be 
$\pm 0.1$~dex.   The resulting intrinsic H$\beta$ luminosity density and
total error for the $z \sim 2$ universe is then
$\log \rho_{{\rm H}\beta} = 39.49 \pm 0.12$ (ergs~s$^{-1}$~Mpc$^{-3}$).  

Our H$\beta$ luminosity function is the first such measurement at
$z \gtrsim 2$.  However, there have been previous estimates of 
the epoch's H$\alpha$ luminosity function from deep, narrow-band surveys in
the infrared.  While \citet{hayes+10} found 
$\log L^*({\rm H}\alpha) = 43.22$ (ergs~s$^{-1}$) 
and a total H$\alpha$ luminosity density of 
$\log \rho_{{\rm H}\alpha} = 39.77$ (ergs~s$^{-1}$~Mpc$^{-3}$) 
in a 56~arcmin$^2$ region of GOODS-S, a much larger ($\sim 2$~deg$^2$) study
by \citet{sobral+13} inferred $\log L^*({\rm H}\alpha) = 42.47$ and
$\log \rho_{{\rm H}\alpha} = 40.01$.  Our best-fit H$\beta$ function, coupled
with an effective logarithmic extinction of $c_{{\rm H}\beta} = 0.5$, a
\citet{cardelli+89} extinction law (with $R_V = 3.1$) and an assumed
intrinsic H$\alpha$/H$\beta$ ratio of 2.86 \citep{pengelly64, AGN3} implies 
observed values for $L^*$ and the H$\alpha$ luminosity density of
$\log L({\rm H}\alpha) = 42.68$ (ergs~s$^{-1}$) and 
$\log \rho_{{\rm H}\alpha} = 39.70$ (ergs~s$^{-1}$~Mpc$^{-3}$).  In
other words, our measurement of $L^*$ is in good agreement with that of 
\citet{sobral+13}, but our estimated luminosity density is more in line 
with that found by \citet{hayes+10}.  It is somewhat surprising that
these three surveys differ by more than 0.3~dex in their determination of 
luminosity density, but given the small volumes involved (5440, 
77,000, and 108,000~Mpc$^3$ for the \citet{hayes+10}, \citet{sobral+13},
and this survey, respectively), and the expected cosmic variance in the
number counts \citep[$\sim 50\%, 15\%$, and 20\%;][]{trenti+08}, 
the result is still reasonable.

We can also test our measurement of the H$\beta$ luminosity function by
converting it into a star formation rate.  The observed H$\beta$ luminosity 
density in our {\sl HST\/} fields is $\log\rho_{{\rm H}\beta} = 38.99 \pm 0.04$
(ergs~s$^{-1}$~Mpc$^{-3}$).  If we again de-redden this number by
$c_{{\rm H}\beta} = 0.5$ and assume an intrinsic H$\alpha$/H$\beta$ ratio of
2.86, then $\log \rho_{{\rm H}\alpha} = 39.95$ (ergs~s$^{-1}$~Mpc$^{-3}$).
The application of the local H$\alpha$ star-formation rate calibration 
\citep{hao+11, murphy+11, kennicutt+12} then yields $\log \rho_{\rm SFR} 
\sim -1.3 \pm 0.1$ ($M_{\odot}$~yr$^{-1}$~Mpc$^{-3}$) where the
error is dominated by the uncertainty in the reddening correction.  This
value is generally consistent with most other determinations of the epoch's
star-formation rate density \citep[e.g.,][]{hopkins04, wilkins+08, bouwens+10},
although, as pointed out above, it is slightly lower than that found
from the H$\alpha$ observations of \citet{sobral+13}.

\subsection{The Ly$\alpha$ Luminosity Function and \fesc}
\label{subsec:LAE_LF}
To compute the volumetric Ly$\alpha$ escape fraction of the
$z \sim 2$ universe, we repeated the above analysis using the Ly$\alpha$
emitting galaxies found in our overlapping survey area.  For the HPS
objects, this involved using the estimates of completeness versus line flux
computed by \citet{hetdex-1} for the 186 separate VIRUS-P pointings covering
the {\sl HST-}grism fields in COSMOS and GOODS-N\null; for the narrow-band data,
the completeness fraction versus monochromatic flux relation was based on the 
artificial star experiments performed by \citet{ciardullo+12} on the region's 
deep Mosaic images.  These functions were then folded into the $1/V_{\rm max}$ 
calculation of equation~(\ref{eq:vmax}) and the luminosity function was fit 
using the maximum-likelihood procedures described by 
equation~(\ref{eq:likelihood}).  

\begin{figure*}[t]
\figurenum{3}
\plotone{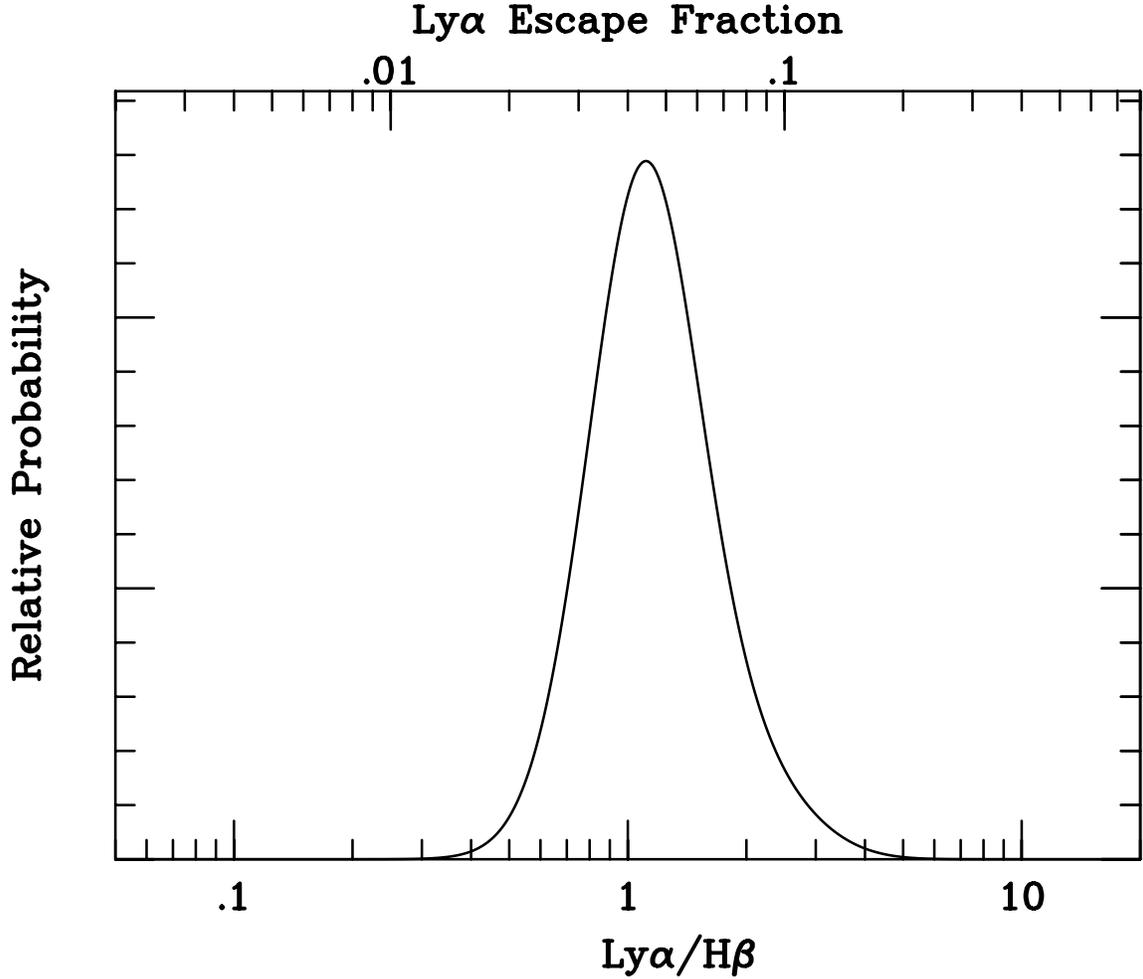}
\caption{Our maximum likelihood solution for the escape fraction of
Ly$\alpha$ photons from our $\sim 100,000$~Mpc$^3$ survey regions at
$1.92 < z < 2.35$.  The plot assumes steady-state star formation for
at least $\sim 0.5$~Gyr, an effective logarithmic H$\beta$ 
extinction of $c_{{\rm H}\beta} = 0.5$, that $\sim 20\%$ of Ly$\alpha$
photons arise from low equivalent width objects, and that Case B 
recombination holds.  Errors in the latter two assumptions would likely
drive the escape fraction to lower values.}
\label{fig:escape_prb}
\end{figure*}

The results are shown in the bottom panels of Figure~\ref{fig:lf} and 
summarized in Table~\ref{tab:lf}.
Because our survey fields contain far fewer LAEs than H$\beta$ galaxies,
the Ly$\alpha$ luminosity function is not as well defined as its H$\beta$
counterpart.   Indeed, only 11 HPS galaxies and 6 narrow-band selected
objects have Ly$\alpha$ detections that are brighter than their frame's 50\%
completeness limit.  As a result, the ``knee'' of the \citet{schechter} 
function is poorly defined and detected only with $\sim 2 \, \sigma$ 
significance.  Nevertheless, the best fit function, with $\log L = 43.60$
(ergs~s$^{-1}$) and $\log \phi^* = -4.45$ (Mpc$^{-3}$), is similar to that 
found by \citet{blanc+11} for the entire sample of 49 non-X-ray-emitting 
HPS LAEs with $1.9 < z < 2.8$.  Our $z \sim 2$ value of $\log L^*$ is 
significantly larger that those found by narrow-band surveys 
\citep{ciardullo+12, gronwall+14}, but, given the vastly larger volume studied 
by the HPS and the increased likelihood of finding rare, exceptionally bright 
objects, this difference is not a serious concern.

Despite the rather large uncertainty associated with $\log L^*$, the
total Ly$\alpha$ luminosity density of the $z \sim 2$ universe, as
defined through equation~(\ref{eq:sch_int}), is very well defined, with
$\log \rho_{{\rm Ly}\alpha} = 39.56 \pm 0.11$~(ergs~s$^{-1}$~Mpc$^{-3}$).
This number is a slight underestimate, since, unlike the {\sl HST\/} grism,
the HPS and the narrow-band surveys do not unambiguously identify every
Ly$\alpha$ source above a flux limit.  To be classified as an 
LAE, a galaxy must also have a rest-frame equivalent width greater than 
20~\AA\null. (Lower equivalent width objects can be confused with foreground 
sources and are usually ignored.)  To correct for these missing objects, one 
must estimate the total amount of emission associated with low-equivalent 
width galaxies.  If the equivalent width distribution of such objects follows 
the exponential function defined by the higher EW LAEs, then our census may be 
missing between 20\% and 30\% of escaping Ly$\alpha$ photons 
\citep{ciardullo+12, gronwall+14}.  Alternatively, if, as suggested by 
\citet{shimasaku+06}, the true distribution of Ly$\alpha$ equivalent widths 
has a lognormal form similar to that associated with local [O~II] emitting 
galaxies \citep[e.g.,][]{blanton-lin}, then the missing Ly$\alpha$ flux would 
be far less.  Since observations of Lyman-break galaxies appear to support the 
former possibility \citep[e.g.,][]{shapley+03, kornei+10}, we use this more 
conservative approach in our calculations and assume that we are missing 
$\sim 20\%$ of the epoch's total Ly$\alpha$ luminosity density.

Figure~\ref{fig:escape_prb} combines our two likelihood functions to show
the total Ly$\alpha$ escape fraction within our $\sim 100,000$~Mpc$^3$ survey
volume.  In the figure, the most probable escape fraction for the
$z \sim 2$ universe is $4.4^{+2.1}_{-1.2}\%$, with the bulk of our statistical
error arising from the uncertainty in the Ly$\alpha$ luminosity function.  We 
must note, however, that the systematic errors associated with our measurement 
may be greater than this.  While our assumptions about Case~B recombination,
the history of $z \sim 2$ star formation, and the population of low
equivalent width Ly$\alpha$ emitters suggest that our measurement of
\fesc\  is an upper limit, the greatest uncertainty is that
associated with internal extinction.  As mentioned above, none of the objects
studied in this program have measured Balmer decrements; instead, we have
estimates of the stellar reddenings derived from the observed slopes of the 
rest-frame UV continua.  These are, at best, indirectly related to the nebular 
extinctions applicable to our problem \citep{calzetti01}.  To produce
Figure~\ref{fig:escape_prb}, we have assumed $c_{{\rm H}\beta} = 0.5$, but
this number likely has an uncertainty of $\pm 0.1$~dex.  This factor adds an 
additional $\sim 25\%$ error onto our estimate, making our systematic 
error term comparable to our statistical uncertainty.

\section{Discussion and Conclusions}
\label{sec:Discussion}
Our estimate of $4.4^{+2.1}_{-1.2}\%$ for the Ly$\alpha$ escape fraction
is consistent with the study by \citet{blanc+11}, who normalized their
$1.9 < z < 2.8$ LAE observations via rest-frame UV measurements of the 
epoch's star formation rate density.  It also agrees with most models for the 
evolution of \fesc\ with redshift \citep{hayes+11, blanc+11, dijkstra+13}.  
But these analyses rely predominantly on indirect measurements:  in fact, the 
only previous study to directly compare Ly$\alpha$ emission with Ly$\alpha$
production is that of \citet{hayes+10}.   By performing dual narrow-band
surveys in Ly$\alpha$ and H$\alpha$, and estimating nebular extinction via
the reddening of the stellar rest-frame UV continuum, \citet{hayes+10}
obtained $5.3 \pm 3.8$\% as the Ly$\alpha$ escape fraction for the 
$z = 2.2$ universe.  Our $1.92 < z < 2.35$ measurements, which also rely
on stellar reddenings, encompass a volume $\sim 20$ times larger than that 
of the \citet{hayes+10} survey, and are therefore far less sensitive to 
issues associated with cosmic variance.  
More importantly, by surveying
this larger volume, we have been able to better define the bright end
of the Ly$\alpha$ and H$\beta$ luminosity functions.  While our data are not as
good as those of \citet{hayes+10} for defining the slope of the faint-end of 
the luminosity function (we adopt $\alpha = -1.6$, rather than fitting for the
variable), we have many more $L > L^*$ galaxies, and, unless the
\citet{schechter} function has a very steep, faint-end slope, 
it is this latter parameter which is most important
for defining the total emission-line flux. 
Of course, our observations
also have the drawback of being based on H$\beta$, rather than H$\alpha$, and
are thus more sensitive to the effects of dust and internal extinction 
than the previous study.  

At present, our measurements of \fesc\ are limited by two factors:  the depth 
of our survey for Ly$\alpha$, and our ability to estimate $c_{{\rm H}\beta}$ 
for individual galaxies.  Both of these issues should improve rather rapidly.  

At the wavelengths considered here, the HPS typically reached a
monochromatic flux limit of $\sim 1.5 \times 10^{-16}$~ergs~cm$^{-2}$~s$^{-1}$,
or $\log L({{\rm Ly}\alpha}) \sim 42.7$~(ergs~s$^{-1}$) at $z \sim 2.2$.   
Unfortunately, as shown in Figure~\ref{fig:ratio}, this is not deep enough
to detect many of the emission-line galaxies found by the {\sl HST\/} grism,
as our median upper limit on the Ly$\alpha$ escape fraction is 
$\sim 6\%$\null.  In other words, our detection threshold lies just above that 
expected from our analysis of the epoch's volumetric escape fraction.  The 
main HETDEX survey, which begins in 2015, is designed to reach 
Ly$\alpha$ flux limits that are a factor of $\sim 3$ fainter than that for
the HPS, i.e., $\sim 0.5 \times 10^{-16}$~ergs~cm$^{-2}$~s$^{-1}$ at 
$\sim 3900$~\AA\null.  This limit should allow us to directly measure \fesc\ 
for the typical star-forming galaxy of the epoch, and enable a search for 
trends with object size, stellar mass, and star formation rate.

Similarly, our ability to measure the intrinsic Balmer line luminosities of the
H$\beta$-selected galaxies is rapidly advancing.  Like most previous studies of
Ly$\alpha$ emission at $2 < z < 3$ \citep[e.g.,][]{gronwall+07, hayes+10, 
blanc+11}, our estimates of nebular extinction are inferred from measurements
of stellar reddening, and, while there may be a relation between the two 
parameters, this unknown does introduce an uncertainty into the calculation.
Fortunately, instruments are now available which allow simultaneous
infrared spectroscopy for large numbers of $z \sim 2$ sources 
\citep[e.g.,][]{flamingos-2, mosfire}, thereby enabling direct measurements
of the galaxies' Balmer decrements.  This capability will remove the most 
important source of systematic error from the analysis.

\acknowledgments
This work was supported via NSF through grant AST 09-26641\null.  The 
Institute for Gravitation and the Cosmos is supported by the Eberly College of 
Science and the Office of the Senior Vice President for Research at the 
Pennsylvania State University.  STScI declined to support this use of the
3D-HST Treasury Program data.

{\it Facilities:} \facility{HST (WFC3)}, \facility{Smith (VIRUS-P)}, 
\facility{Blanco (Mosaic)}

\clearpage

\LongTables
\begin{deluxetable*}{cccccc}
\tablewidth{0pt}
\tabletypesize{\scriptsize}
\tablecaption{H$\beta$ and Ly$\alpha$ Galaxy Measurements}
\tablehead{
&&&&\multicolumn{2}{c}{$\log$ Flux (ergs~cm$^{-2}$~s$^{-1}$)} \\
\colhead{$\alpha(2000)$} &\colhead{$\delta(2000)$} &\colhead{$z$} 
&\colhead{$\beta$} &\colhead{H$\beta$} &\colhead{Ly$\alpha$} }
\startdata
\multicolumn{3}{l}{COSMOS} \\
10:00:35.54  &$+$02:13:03.0  &1.934  &$-1.25 \pm 0.06$  &$-16.57 \pm 0.11$  &$< -15.88$ \\
10:00:36.67  &$+$02:13:07.7  &2.092  &$-1.55 \pm 0.12$  &$-17.18 \pm 0.45$  &$< -16.12$ \\
10:00:33.96  &$$+02:13:16.0  &2.230  &$-2.30 \pm 0.05$  &\dots              &$-15.55_{-0.07}^{+0.12}$ \\
10:00:17.30  &$+$02:19:26.5  &2.084  &$-1.54 \pm 0.06$  &$-17.27 \pm 0.36$  &$< -15.87$ \\
10:00:15.73  &$+$02:20:28.1  &1.975  &$-1.64 \pm 0.15$  &$-16.78 \pm 0.08$  &$< -15.81$ \\
10:00:47.16  &$+$02:17:44.5  &2.025  &$-1.62 \pm 0.02$  &$-16.09 \pm 0.03$  &$< -15.82$ \\
10:00:41.65  &$+$02:18:00.2  &2.095  &$-1.39 \pm 0.03$  &$-16.84 \pm 0.16$  &$< -15.98$ \\
10:00:40.82  &$+$02:18:22.9  &2.070  &$-1.63 \pm 0.01$  &\dots              &$-15.59_{-0.17}^{+0.21}$ \\
10:00:42.91  &$+$02:18:25.6  &2.096  &$-1.90 \pm 0.04$  &$-16.73 \pm 0.09$  &$< -15.91$ \\
10:00:38.64  &$+$02:18:36.3  &1.927  &$-1.79 \pm 0.08$  &$-16.36 \pm 0.08$  &$< -15.79$ \\
10:00:42.21  &$+$02:18:48.5  &2.290  &$-1.85 \pm 0.08$  &$-16.59 \pm 0.07$  &$< -16.16$ \\
10:00:18.68  &$+$02:14:59.9  &2.310  &$-0.33 \pm 0.70$  &\dots              &$-15.41_{-0.10}^{+0.13}$ \\
10:00:21.92  &$+$02:15:40.0  &2.093  &$-2.46 \pm 0.11$  &$-16.71 \pm 0.08$  &$< -16.15$ \\
10:00:20.69  &$+$02:12:53.4  &2.163  &$-1.55 \pm 0.11$  &$-16.65 \pm 0.08$  &$< -16.21$ \\
10:00:23.79  &$+$02:13:10.4  &2.107  &$-2.50 \pm 0.24$  &$< -17.06$         &$-15.99_{-0.17}^{+0.22}$ \\
10:00:17.22  &$+$02:13:38.0  &2.105  &$-1.67 \pm 0.09$  &$-16.89 \pm 0.14$  &$< -16.16$ \\
10:00:19.19  &$+$02:14:06.6  &2.103  &$-1.86 \pm 0.05$  &$-16.61 \pm 0.09$  &$< -16.31$ \\
10:00:23.61  &$+$02:15:57.4  &2.088  &$-1.57 \pm 0.05$  &$-16.43 \pm 0.06$  &$< -16.17$ \\
10:00:29.58  &$+$02:17:02.8  &1.921  &$-1.86 \pm 0.03$  &$-16.35 \pm 0.05$  &$< -15.45$ \\
10:00:28.64  &$+$02:17:48.7  &2.093  &$-2.49 \pm 0.04$  &\dots              &$-14.84_{-0.03}^{+0.07}$ \\
10:00:27.24  &$+$02:17:31.6  &2.283  &$-2.08 \pm 0.07$  &$-16.70 \pm 0.20$  &$-15.35_{-0.11}^{+0.13}$ \\
10:00:22.88  &$+$02:17:14.0  &2.220  &$-1.71 \pm 0.10$  &$-16.82 \pm 0.11$  &$< -16.21$ \\
10:00:26.61  &$+$02:17:14.5  &2.224  &$-2.65 \pm 0.12$  &$-17.13 \pm 0.22$  &$< -16.18$ \\
10:00:24.22  &$+$02:14:11.7  &2.104  &$-1.32 \pm 0.05$  &$-16.81 \pm 0.14$  &$< -16.26$ \\
10:00:25.45  &$+$02:14:27.8  &2.166  &$-2.17 \pm 0.16$  &$-16.67 \pm 0.09$  &$< -16.18$ \\
10:00:23.41  &$+$02:14:32.3  &2.097  &$-1.79 \pm 0.14$  &$-16.74 \pm 0.08$  &$< -16.10$ \\
10:00:23.23  &$+$02:12:30.7  &2.225  &$-2.45 \pm 0.11$  &$-16.40 \pm 0.08$  &$< -16.20$ \\
10:00:39.12  &$+$02:14:51.0  &2.125  &$-1.51 \pm 0.04$  &$-16.49 \pm 0.06$  &$< -15.73$ \\
10:00:32.33  &$+$02:14:53.0  &1.975  &$-2.37 \pm 0.06$  &$-16.37 \pm 0.05$  &$< -15.58$ \\
10:00:32.60  &$+$02:15:35.9  &2.162  &$-1.87 \pm 0.05$  &$-16.45 \pm 0.06$  &$< -15.83$ \\
10:00:34.08  &$+$02:15:54.4  &2.199  &$-1.81 \pm 0.13$  &$-16.99 \pm 0.17$  &$< -15.79$ \\
10:00:44.93  &$+$02:15:53.2  &2.093  &$-2.38 \pm 0.13$  &$-16.84 \pm 0.08$  &$< -15.67$ \\
10:00:37.02  &$+$02:17:47.5  &1.940  &$-2.11 \pm 0.14$  &$-16.96 \pm 0.11$  &$< -15.72$ \\
10:00:29.81  &$+$02:18:49.2  &2.200  &$-1.69 \pm 0.05$  &\dots              &$-15.38_{-0.06}^{+0.06}$ \\
10:00:19.55  &$+$02:17:56.1  &2.052  &$-1.84 \pm 0.07$  &$-16.97 \pm 0.31$  &$< -15.85$ \\
\multicolumn{3}{l}{GOODS-N} \\
12:36:15.00  &$+$62:13:29.9  &1.998  &$-1.97 \pm 0.20$  &$-16.66 \pm 0.06$  &$< -15.72$ \\    
12:36:20.46  &$+$62:14:52.2  &1.999  &$-2.26 \pm 0.38$  &$-17.21 \pm 0.15$  &$< -15.29$ \\
12:36:15.86  &$+$62:13:25.7  &2.086  &$-2.20 \pm 0.30$  &$-17.58 \pm 0.37$  &$< -15.90$ \\
12:36:18.78  &$+$62:10:37.3  &2.264  &$-1.73 \pm 0.12$  &$-16.55 \pm 0.06$  &$< -16.11$ \\
12:36:20.07  &$+$62:11:12.4  &2.004  &$-1.64 \pm 0.11$  &$-16.76 \pm 0.09$  &$< -15.75$ \\
12:36:13.33  &$+$62:11:45.2  &2.258  &$-1.74 \pm 0.25$  &$-16.93 \pm 0.11$  &$< -15.93$ \\
12:36:24.96  &$+$62:12:23.6  &2.216  &$-1.56 \pm 0.18$  &$-17.04 \pm 0.16$  &$< -16.10$ \\
12:36:40.61  &$+$62:13:10.9  &2.051  &$-1.81 \pm 0.25$  &$-16.89 \pm 0.13$  &$< -16.09$ \\
12:36:35.42  &$+$62:14:37.6  &2.006  &$-1.68 \pm 0.06$  &$-16.75 \pm 0.41$  &$< -16.07$ \\
12:36:42.09  &$+$62:13:31.4  &2.018  &$+0.52 \pm 0.51$  &$-17.25 \pm 0.27$  &$< -16.05$ \\
12:36:44.90  &$+$62:13:35.6  &2.234  &$-1.85 \pm 0.43$  &$-17.14 \pm 0.12$  &$< -16.35$ \\
12:36:50.10  &$+$62:14:01.2  &2.235  &$-1.49 \pm 0.10$  &\dots              &$-15.61_{-0.06}^{+0.09}$ \\
12:36:44.12  &$+$62:14:01.9  &2.273  &$-0.12 \pm 1.25$  &\dots              &$-15.69_{-0.21}^{+0.16}$ \\
12:36:47.46  &$+$62:15:03.5  &2.083  &$-2.46 \pm 0.21$  &$-16.69 \pm 0.06$  &$-15.94_{-0.08}^{+0.08}$ \\
12:36:53.79  &$+$62:15:21.7  &2.022  &$-2.54 \pm 0.26$  &$-17.22 \pm 0.24$  &$< -15.83$ \\
12:36:55.60  &$+$62:14:50.7  &1.975  &$-1.85 \pm 0.21$  &$-17.11 \pm 0.13$  &$< -16.21$ \\
12:36:41.26  &$+$62:11:15.6  &2.062  &$-1.96 \pm 0.12$  &$-16.38 \pm 0.05$  &$< -15.84$ \\
12:36:55.06  &$+$62:13:47.1  &2.233  &$-2.10 \pm 0.10$  &$-16.57 \pm 0.05$  &$< -16.33$ \\
12:36:54.19  &$+$62:13:35.9  &2.263  &$-2.59 \pm 0.59$  &$-16.86 \pm 0.08$  &$-15.92_{-0.11}^{+0.15}$ \\
12:37:04.33  &$+$62:14:46.2  &2.220  &$-1.76 \pm 0.10$  &$-16.33 \pm 0.07$  &$-15.11_{-0.04}^{+0.05}$ \\
12:37:11.78  &$+$62:13:38.9  &1.914  &$-2.08 \pm 0.17$  &$-16.71 \pm 0.06$  &$< -15.79$ \\
12:37:11.20  &$+$62:14:31.5  &2.191  &$-2.24 \pm 0.24$  &$-17.42 \pm 0.27$  &$< -16.27$ \\
12:37:02.01  &$+$62:14:19.1  &2.291  &$-2.21 \pm 0.53$  &$-17.09 \pm 0.11$  &$< -16.16$ \\
12:36:53.32  &$+$62:10:35.9  &1.975  &$-1.69 \pm 0.09$  &$-16.78 \pm 0.11$  &$< -16.04$ \\
12:37:02.72  &$+$62:10:10.8  &1.989  &$-1.66 \pm 0.16$  &$-16.51 \pm 0.04$  &$< -15.52$ \\
12:37:11.00  &$+$62:11:40.1  &2.270  &$-2.10 \pm 0.15$  &\dots              &$-15.86_{-0.28}^{+0.20}$ \\
12:37:07.10  &$+$62:11:52.6  &2.275  &$-1.89 \pm 0.11$  &$-17.22 \pm 0.20$  &$< -16.10$ \\
12:37:10.42  &$+$62:10:35.5  &2.161  &$-2.02 \pm 0.58$  &$-17.43 \pm 0.32$  &$< -15.85$ \\
\multicolumn{3}{l}{GOODS-S} \\
 3:33:00.61  &$-$27:40:27.0  &2.030  &$-1.55 \pm 2.73$  &$< -17.12$         &$-16.69_{-0.09}^{+0.09}$ \\
 3:32:40.52  &$-$27:49:32.5  &2.042  &$-2.06 \pm 1.22$  &$-17.06 \pm 0.20$  &$-16.49_{-0.09}^{+0.09}$ \\
 3:32:34.44  &$-$27:47:42.8  &2.030  &$-2.00 \pm 0.13$  &$-16.73 \pm 0.05$  &\dots \\
 3:32:41.55  &$-$27:48:24.4  &2.080  &$-2.25 \pm 0.44$  &$-16.88 \pm 0.25$  &$-16.50_{-0.07}^{+0.07}$ \\
 3:32:42.19  &$-$27:48:59.6  &2.079  &$-1.94 \pm 0.26$  &$-17.54 \pm 0.49$  &$-15.99_{-0.02}^{+0.02}$ \\
 3:32:28.85  &$-$27:52:20.1  &2.039  &$-1.47 \pm 0.18$  &$-16.41 \pm 0.05$  &\dots \\
 3:32:13.76  &$-$27:43:00.5  &2.070  &$-2.50 \pm 0.13$  &$-16.83 \pm 0.27$  &$-16.13_{-0.03}^{+0.03}$ \\
 3:32:45.54  &$-$27:53:43.3  &2.059  &$-1.55 \pm 0.22$  &$-17.17 \pm 0.26$  &$-16.53_{-0.08}^{+0.08}$ \\
 3:32:26.83  &$-$27:46:01.8  &2.079  &$-1.15 \pm 0.15$  &$-16.75 \pm 0.07$  &\dots \\
 3:32:25.82  &$-27$:46:09.3  &2.076  &$-1.56 \pm 0.30$  &$-17.02 \pm 0.13$  &\dots \\ 
 3:32:11.45  &$-$27:50:26.7  &2.068  &$-2.37 \pm 0.18$  &$-16.29 \pm 0.04$  &$-15.97_{-0.02}^{+0.02}$ \\
 3:32:18.12  &$-$27:49:41.9  &2.034  &$-1.77 \pm 0.08$  &$-16.90 \pm 0.19$  &\dots \\
 3:32:59.11  &$-$27:53:20.7  &2.030  &$-0.03 \pm 0.09$  &$-16.55 \pm 0.18$  &$-16.02_{-0.02}^{+0.02}$ \\
 3:32:04.00  &$-$27:43:51.6  &2.044  &$-2.02 \pm 0.58$  &\dots              &$-16.65_{-0.10}^{+0.10}$ \\
 3:32:08.66  &$-$27:42:50.2  &2.075  &$-1.85 \pm 0.12$  &$-16.66 \pm 0.16$  &\dots \\
 3:32:07.84  &$-$27:42:27.2  &2.049  &$-1.22 \pm 0.44$  &$-16.61 \pm 0.06$  &\dots \\
 3:32:09.60  &$-$27:47:39.9  &2.040  &$-1.53 \pm 0.21$  &$-17.04 \pm 0.12$  &\dots \\
 3:32:21.40  &$-$27:51:26.2  &2.037  &$-1.14 \pm 0.20$  &$-16.52 \pm 0.08$  &\dots \\
 3:32:46.15  &$-$27:49:22.6  &2.032  &$-2.08 \pm 0.10$  &$-17.01 \pm 0.14$  &\dots \\
 3:32:43.46  &$-$27:43:36.5  &2.078  &$-0.51 \pm 0.23$  &$-16.90 \pm 0.13$  &\dots \\
 3:32:23.24  &$-$27:42:31.6  &2.078  &$-1.66 \pm 0.16$  &$-16.64 \pm 0.04$  &\dots \\
 3:32:35.73  &$-$27:46:39.0  &2.073  &$-0.50 \pm 0.60$  &$-16.99 \pm 0.07$  &\dots \\
 3:32:35.60  &$-$27:47:45.4  &2.030  &$-1.48 \pm 0.49$  &$-17.20 \pm 0.10$  &\dots \\
3:32:40.19  &$-$27:46:54.6  &2.069  &$-0.64 \pm 0.92$  &$-17.35 \pm 0.14$  &$-16.36_{-0.05}^{+0.05}$  
\enddata
\label{tab:galaxies}
\end{deluxetable*}
\clearpage

\begin{deluxetable}{lccc}
\tablewidth{0pt}
\tablecaption{Best-Fit Schechter Function Parameters}
\tablehead{
           &\colhead{Entire HST}  &\multicolumn{2}{c}{HST/HPS + NB}  \\ 
           &                         &\multicolumn{2}{c}{overlap} \\
\colhead{Parameter} &\colhead{H$\beta$}  &\colhead{H$\beta$} 
&\colhead{Ly$\alpha$} }
\startdata
Area (arcmin$^2$)   &275   &157   &157 \\
Maximum Volume (Mpc$^3$)  &408,000 &103,000 &103,000 \\
Number of Galaxies  &97 &43 &17 \\
$c_{{\rm H}\beta}$ (median) &0.37 &0.37 &0.30 \\
$c_{{\rm H}\beta}$ (effective) &0.55 &0.44 &0.89 \\ 
Fixed $\alpha$ &$-1.6$ &$-1.6$ &$-1.6$ \\
\noalign{\vskip6pt}
\multicolumn{4}{l}{Fitted Quantities Prior to Extinction Correction} \\ 
\noalign{\vskip6pt}
$\log L^*$ (ergs~s$^{-1}$) &$42.07 \pm 0.09$ &$42.07 \pm 0.13$ 
&$43.60^{+0.54}_{-0.23}$ \\
$\phi_{\rm tot}$ ($\log L < 41.5$) (Mpc$^{-3}$) 
&$-3.28 \pm 0.04$ &$-3.18 \pm 0.07$ &$-3.01 \pm 0.12$ \\
$\log \rho$ (ergs~s$^{-1}$~Mpc$^{-3}$) &$38.99 \pm 0.04$ &$39.09 \pm 0.06$ 
&$39.56 \pm 0.11$ \\
Best $\phi^*$ (Mpc$^3)$ &$-3.43$ &$-3.32$ &$-4.44$ \\
\enddata
\label{tab:lf}
\end{deluxetable}

\clearpage

\begin{thebibliography}{apj}

\bibitem[Adams et al.~(2011)]{hetdex-1} Adams, J.J., Blanc, G.A.,
Hill, G.J., et al.~\ 2011, \apjs, 192, 5 

\bibitem[Alexander et al.~(2003)]{alexander+03} Alexander, D.M., 
Bauer, F.E., Brandt, W.N., et al.~\  2003, \aj, 126, 539 

\bibitem[An et al.~(2014)]{an+14} An, F.X., Zheng, X.Z., 
Wang, W.-H., et al.~\  2014, \apj, 784, 152 

\bibitem[Blanc et al.~(2011)]{blanc+11} Blanc, G.A., Adams, J.J., Gebhardt, K.,
et al.~\  2011, \apj, 736, 31

\bibitem[Blanton \& Lin(2000)]{blanton-lin} Blanton, M., \& Lin, H. 2000,
\apjl, 543, L125

\bibitem[Bouwens et al.~(2010)]{bouwens+10} Bouwens, R.J., 
Illingworth, G.D., Oesch, P.A., et al.~\  2010, \apjl, 709, L133 

\bibitem[Brammer et al.~(2012)]{3DHST} Brammer, G.B., van Dokkum, P.G., 
Franx, M., et al.~\ 2012, \apjs, 200, 13 

\bibitem[Buat et al.~(2011)]{buat+11} Buat, V., Giovannoli, E., Heinis, S.,
et al.~\ 2011, \aap, 553, A93

\bibitem[Calzetti(2001)]{calzetti01} Calzetti, D. 2001, \pasp, 113, 1449 

\bibitem[Cardelli et al.~(1989)]{cardelli+89} Cardelli, J.A., 
Clayton, G.C., \& Mathis, J.S. 1989, \apj, 345, 245 

\bibitem[Cassata et al.~(2011)]{cassata+11} Cassata, P., Le F\`evre, O., 
Garilli, B., et al.~\  2011, \aap, 525, A143 

\bibitem[Charlot \& Fall(2000)]{cha-fall00} Charlot, S., \& Fall, S.M. 2000, 
\apj, 539, 718 

\bibitem[Chen et al.~(2007)]{chen+07} Chen, H.-W., Prochaska, J.X., \& 
Gnedin, N.Y. 2007, \apjl, 667, L125

\bibitem[Ciardullo et al.~(2012)]{ciardullo+12} Ciardullo, R.,
Gronwall, C., Wolf, C., et al.~\ 2012, \apj, 744, 110

\bibitem[Ciardullo et al.~(2013)]{ciardullo+13} Ciardullo, R., 
Gronwall, C., Adams, J.J., et al.~\ 2013, \apj, 769, 83 

\bibitem[Cohen(1991)]{cohen91} Cohen, A.C. 1991, Truncated and Censored 
Samples: Theory and Applications (New York:  Marcel Dekker)

\bibitem[Colbert et al.~(2013)]{colbert+13} Colbert, J.W., 
Teplitz, H., Atek, H., et al.\ 2013, \apj, 779, 34 

\bibitem[Cowie et al.~(2010)]{cowie+10} Cowie, L.L., Barger, A.J., \&
Hu, E.M. 2010, \apj, 711, 928

\bibitem[Deharveng et al.~(2008)]{deharveng+08} Deharveng, J.-M., Small, T.,
Barlow, T.A., et al.~\ 2008, \apj, 680, 1072

\bibitem[Dijkstra \& Jeeson-Daniel(2013)]{dijkstra+13} Dijkstra, M., \& 
Jeeson-Daniel, A. 2013, \mnras, 435, 3333

\bibitem[Dom{\'{\i}}nguez et al.~(2013)]{dominguez+13}
Dom{\'{\i}}nguez, A., Siana, B., Henry, A.L., et al.~\  2013, \apj, 763, 145 

\bibitem[Eikenberry et al.~(2012)]{flamingos-2} Eikenberry, S., 
Bandyopadhyay, R., Bennett, J.G., et al.~\  2012, \procspie, 8446

\bibitem[Elvis et al.~(2009)]{elvis+09} Elvis, M., Civano, F., 
Vignali, C., et al.~\  2009, \apjs, 184, 158

\bibitem[Feigelson \& Babu(2012)]{feigelson-babu} Feigelson, E.D., \& 
Babu, G.J. 2012, Modern Statistical Methods for Astronomy with R Applications 
(Cambridge: Cambridge University Press)

\bibitem[Finkelstein et al.~(2008)]{finkelstein+08} Finkelstein, S.L., 
Rhoads, J.E., Malhotra, S., Grogin, N., \& Wang, J. 2008, \apj, 678, 655 

\bibitem[Finkelstein et al.~(2009)]{finkelstein+09} Finkelstein, S.L.,       
Rhoads, J.E., Malhotra, S., \& Grogin, N. 2009, \apj, 691, 465   

\bibitem[Finkelstein et al.~(2011)]{finkelstein+11} Finkelstein, S.L.,       
Cohen, S.H., Moustakas, J., et al.~\  2011, \apj, 733, 117  

\bibitem[Foreman-Mackey et al.~(2013)]{emcee} Foreman-Mackey, D.,
Hogg, D.W., Lang, D., \& Goodman, J. 2013, \pasp, 125, 306 

\bibitem[F\"orster Schreiber et al.~(2009)]{forster+09} 
F\"orster Schreiber, N.M., Genzel, R., Bouch\'e, N., et al.~\  2009, 
\apj, 706, 1364 

\bibitem[Garn \& Best(2010)]{garn-best} Garn, T., \& Best, P.N. 2010,
\mnras, 409, 421

\bibitem[Gebhardt et al.~(2014)]{gebhardt+14} Gebhardt, H., Zeimann, G.R.,
Ciardullo, R., \& Gronwall, C. 2014, \apj, in preparation

\bibitem[Giavalisco et al.~(2004)]{GOODS} Giavalisco, M., Ferguson, H.C., 
Koekemoer, A.M., et al.~\  2004, \apjl, 600, L93 

\bibitem[Gronwall et al.~(2007)]{gronwall+07} Gronwall, C., Ciardullo, R.,
Hickey, T., et al.~\ 2007, \apj, 667, 79

\bibitem[Gronwall et al.~(2014)]{gronwall+14} Gronwall, C., Ciardullo, R.,
Matkovi\'c, A., et al.~\ 2014, in preparation

\bibitem[Guaita et al.~(2010)]{guaita+10} Guaita, L., Gawiser, E., 
Padilla, N., et al.~\ 2010, \apj, 714, 255 

\bibitem[Guti\'errez \& Beckman(2010)]{gutierrez+10} Guti\'errez, L., \& 
Beckman, J.E. 2010, \apjl, 710, L44

\bibitem[Hagen et al.~(2014a)]{hagen+14a} Hagen, A., Ciardullo, R., 
Gronwall, C., et al.~\ 2014a, \apj, 786, 59 

\bibitem[Hagen et al.~(2014b)]{hagen+14b} Hagen, A., Zeimann, G.R., 
Ciardullo, R., et al.~\ 2014b, in preparation

\bibitem[Hansen \& Oh(2006)]{hansen+06} Hansen, M., \& Oh, S.P. 2006, 
\mnras, 367, 979

\bibitem[Hao et al.~(2011)]{hao+11} Hao, C.N., Kennicutt, R.C., Johnson, B.D., 
et al.~\ 2011, ApJ, 741, 124

\bibitem[Hayes et al.~(2010)]{hayes+10} Hayes, M., {\"O}stlin, G., Schaerer, D.,
et al.~\ 2010, \nat, 464, 562

\bibitem[Hayes et al.~(2011)]{hayes+11} Hayes, M., Schaerer, D.,
{\"O}stlin, G., et al.~\ 2011, \apj, 730, 8

\bibitem[Hill et al.~(2008)]{VIRUS} Hill, G.J., MacQueen, P.J., Smith, M.P.,
et al.~\ 2008, \procspie, 7014, 231

\bibitem[Hinshaw et al.~(2013)]{hinshaw+13} Hinshaw, G., Larson, 
D., Komatsu, E., et al.~\  2013, \apjs, 208, 19 

\bibitem[Holden et al.~(2014)]{holden+14} Holden, B.P., Oesch, P.A.,
Gonzalez, V.G., et al.~\  2014, submitted to ApJ (arXiv:1401.5490)

\bibitem[Hopkins(2004)]{hopkins04} Hopkins, A.M. 2004, \apj, 615, 209 

\bibitem[Huchra \& Sargent(1973)]{huchra73} Huchra, J., \& Sargent, W.L.W.
1973, \apj, 186, 433

\bibitem[Hunter \& Hoffman(1999)]{hunter+99} Hunter, D.A., \& Hoffman, L. 
1999, \aj, 117, 2789 

\bibitem[Iwata et al.~(2009)]{iwata+09} Iwata, I., Inoue, A.K., Matsuda, Y., 
et al.~\  2009, \apj, 692, 1287 

\bibitem[Kennicutt \& Evans(2012)]{kennicutt+12} Kennicutt, R.C., \&
Evans, N.J. 2012, \araa, 50, 531

\bibitem[Kornei et al.~(2010)]{kornei+10} Kornei, K.A., Shapley, A.E.,
Erb, D.K., et al.~\  2010, \apj, 711, 693 

\bibitem[Kriek \& Conroy(2013)]{kriek+13} Kriek, M., \& Conroy, C. 2013, 
\apjl, 775, L16 

\bibitem[Lee \& Wang(2003)]{lee-wang03} Lee, E.T., \& Wang, J. 2003, 
Statistical Methods for Survival Data Analysis, 3rd Edition
(New York: Wiley-Interscience)

\bibitem[Ly et al.~(2011)]{ly+11} Ly, C., Lee, J.C., Dale, D.A., et al.~\ 
2011, \apj, 726, 109 

\bibitem[Mannucci et al.~(2009)]{mannucci+09} Mannucci, F., Cresci, 
G., Maiolino, R., et al.~\  2009, \mnras, 398, 1915 

\bibitem[McLean et al.~(2012)]{mosfire} McLean, I.S., Steidel, C.C.,
Epps, H.W., et al.~\ 2012, \procspie, 8446 

\bibitem[Moustakas et al.~(2006)]{moustakas+06} Moustakas, J., 
Kennicutt, R.C., Jr., \& Tremonti, C.A. 2006, \apj, 642, 775 

\bibitem[Murphy et al.~(2011)]{murphy+11} Murphy, E.J., Condon, J.J., 
Schinnerer, E., et al.~\ 2011, ApJ, 737, 67

\bibitem[Nakajima et al.~(2012)]{nakajima+12} Nakajima, K., Ouchi,         
M., Shimasaku, K., et al.~\  2012, \apj, 745, 12  

\bibitem[Neufeld(1991)]{neufeld91} Neufeld, D.A. 1991, \apjl, 370, L85

\bibitem[Nilsson et al.~(2009)]{nilsson+09} Nilsson, K.K., Tapken, C.,
M\"oller, P.,  et al.~\ 2009, \aap, 498, 13

\bibitem[Osterbrock \& Ferland(2006)]{AGN3} Osterbrock, D.E., \& Ferland, G.J. 
2006, Astrophysics of Gaseous Nebulae and Active Galactic Nuclei, 2nd.~ed.~by 
D.E. Osterbrock \& G.J. Ferland (Sausalito, CA: University Science Books)

\bibitem[Ouchi et al.~(2008)]{ouchi+08} Ouchi, M.,  Shimasaku, K., Akiyama, M.,
et al.~\ \apjs, 176, 301

\bibitem[Partridge \& Peebles(1967)]{partridge+67} Partridge, R.B., \&
Peebles, P.J.E. 1967, \apj, 147, 868

\bibitem[Pengelly(1964)]{pengelly64} Pengelly, R.M. 1964, \mnras, 127, 145 

\bibitem[Price et al.~(2014)]{price+14} Price, S.H., Kriek, M., 
Brammer, G.B., et al.~\  2014, \apj, 788, 86 

\bibitem[Schaerer et al.~(2011)]{schaerer+11} Schaerer, D., Hayes, M.,
Verhamme, A., \& Teyssier, R. 2011, \aap, 531, A12

\bibitem[Schechter(1976)]{schechter} Schechter, P. 1976, \apj, 203, 297 

\bibitem[Schmidt(1968)]{schmidt68} Schmidt, M. 1968, \apj, 151, 393

\bibitem[Scoville et al.~(2007)]{COSMOS} Scoville, N., Aussel, H., 
Brusa, M., et al.~\  2007, \apjs, 172, 1 

\bibitem[Shapley et al.~(2003)]{shapley+03} Shapley, A.E., 
Steidel, C.C., Pettini, M., \& Adelberger, K.L. 2003, \apj, 588, 65 

\bibitem[Shimasaku et al.~(2006)]{shimasaku+06} Shimasaku, K., 
Kashikawa, N., Doi, M., et al.~\  2006, \pasj, 58, 313 

\bibitem[Skelton et al.~(2014)]{skelton+14} Skelton, R.E., 
Whitaker, K.E., Momcheva, I.G., et al.~\ 2014, \apjs, in press (arXiv:1403.3689)

\bibitem[Sobral et al.~(2013)]{sobral+13} Sobral, D., Smail, I., 
Best, P.N., et al.~\  2013, \mnras, 428, 1128 

\bibitem[Trenti \& Stiavelli(2008)]{trenti+08} Trenti, M., \& Stiavelli, M. 
2008, \apj, 676, 767 

\bibitem[van der Wel et al.~(2014)]{vanderwel+14} van der Wel, A., 
Franx, M., van Dokkum, P.G., et al.~\  2014, \apj, 788, 28

\bibitem[Vanzella et al.~(2010)]{vanzella+10} Vanzella, E., Giavalisco, M., 
Inoue, A.K., et al.~\  2010, \apj, 725, 1011 

\bibitem[Verhamme et al.~(2006)]{verhamme+06} Verhamme, A., Schaerer, D., \&
Maselli, A. 2006, \aap, 460, 397

\bibitem[Wardlow et al.~(2014)]{wardlow+14} Wardlow, J.L., 
Malhotra, S., Zheng, Z., et al.~\ 2014, \apj, 787, 9 

\bibitem[Weiner \& the AGHAST Team(2014)]{weiner+14} Weiner, B.J., \& AGHAST Team 
2014, \baas, 223, \#227.07

\bibitem[Wilkins et al.~(2008)]{wilkins+08} Wilkins, S.M., 
Trentham, N., \& Hopkins, A.M. 2008, \mnras, 385, 687 

\bibitem[Wold et al.~(2014)]{wold+14} Wold, I.G.B., Barger, A.J.,
\& Cowie, L.L. 2014, \apj, 783, 119 

\bibitem[Wuyts et al.~(2013)]{wuyts+13} Wuyts, S., F\"orster 
Schreiber, N.M., Nelson, E.J., et al.~\  2013, \apj, 779, 135 

\bibitem[Xue et al.~(2011)]{xue+11} Xue, Y.Q., Luo, B., 
Brandt, W.N., et al.~\  2011, \apjs, 195, 10 

\bibitem[Zeimann et al.~(2014)]{zeimann-1} Zeimann, G.R., Ciardullo, R.,
Gebhardt, H., \& Gronwall, C. 2014, \apj, 790, 113

\bibitem[Zheng et al.~(2012)]{zheng+12} Zheng, Z.-Y., Malhotra, S.,
Wang, J.-X., et al.~\ 2012, \apj, 746, 28

\end{thebibliography}
\end{document}